\documentclass[a4paper,fleqn]{cas-sc}

\def\tsc#1{\csdef{#1}{\textsc{\lowercase{#1}}\xspace}}
\tsc{WGM}
\tsc{QE}

\newtheorem{theorem}{Theorem}
\newdefinition{rmk}{Remark}
\newproof{pf}{Proof}
\usepackage{graphicx}%
\usepackage{amsmath,amssymb,amsfonts}%

\usepackage{mathrsfs}%

\usepackage{xcolor}%
\usepackage{textcomp}%
\usepackage{colonequals}

\newcommand{\bi}{\begin{itemize}}
\newcommand{\ei}{\end{itemize}}
\newcommand{\p}{\partial}
\newcommand{\mc}{\mathcal}
\newcommand{\R}{\mathbb R}

\def\be{\begin{equation}}
\def\ee{\end{equation}}
\def\bn{\begin{enumerate}}
\def\en{\end{enumerate}}

\newcommand{\bea}{\begin{eqnarray}}
\newcommand{\eea}{\end{eqnarray}}
\newcommand{\ba}{\begin{array}}
\newcommand{\ea}{\end{array}}
\newcommand{\bl}{\begin{align}}
\newcommand{\el}{\end{align}}

\newcommand{\ist}{$1^\mathrm{st}~$}

\newcommand{\al}{\alpha}

\newcommand{\lam}{\lambda}

\newcommand{\ep}{\varepsilon}

\newcommand{\bs}{\begin{subequations}}
\newcommand{\es}{\end{subequations} \noindent}

\renewcommand{\Re}{\mathop{\rm Re}\nolimits}

\DeclareMathOperator{\csch}{cosech}
\DeclareMathOperator{\diag}{diag}

\newtheorem{lemma}{Lemma}

\newtheorem{remark}{Remark}%
\newtheorem{corollary}{Corollary}

\numberwithin{remark}{section}
\begin{document}
\let\WriteBookmarks\relax
\def\floatpagepagefraction{1}
\def\textpagefraction{.001}

\shorttitle{A multipurpose system of ODE with exponential
nonlinearity}

\shortauthors{P.P. Goldstein}

\title [mode = title]{A multipurpose system of ordinary differential equations with exponential
nonlinearity}



%

\author[1]{Piotr P. Goldstein}[orcid=0000-0002-0236-5332]

\cormark[1]

\fnmark[1]

\ead{piotr.goldstein@ncbj.gov.pl}



\affiliation[1]{organization={National Centre for Nuclear
Research},
            addressline={Pasteur St. 7},
            city={Warsaw},
          citysep={}, 
            postcode={02-093},
            state={},
            country={Poland}}



\ead{}







\begin{abstract}
A system of ordinary differential equations (ODE)
$\ddot{y}_k=\sum_{l=1}^n M_{kl}\exp y_l,~k=1,...,n$, where $M$ is
an $n\times n$ nonsingular symmetric matrix, is discussed in
detail. A two-parameter family of exact solutions is found and
proved to be unstable for general $n$. A result of Adler and
Moerbeke is shown to apply to our systems; it states that the only
completely integrable cases of this system are equivalent to the
Toda lattice with included non-nearest neighbour interactions. A
few interesting applications are shown for some special $n$'s.
\end{abstract}


\begin{highlights}
\item
Finding a system of ODE which describes several distinct physical
systems.
\item
Finding a family of exact solution of that system for arbitrary
number of equations, and proving its instability to small
perturbations.
\item
Demonstrating that the only integrable case of this system for
arbitrary number of equations is equivalent to the Toda chain.
\end{highlights}


\maketitle

\section{Introduction}  \label{1}
We consider a system of ordinary differential equations (ODE)
\be\label{main}
\left(\ba{c}y_1\\ y_2\\
...\\ y_n\ea\right)^{\scriptscriptstyle\bullet\bullet}
=M\cdot\left(\ba{c} \exp{\,y_1}\\ \exp{\,y_2}\\
...\\ \exp{\,y_n}\ea\right),
\ee
where $M$ is a nonsingular constant symmetric real $n\times n$
matrix, while the dots in the superscripts or over symbols denote
differentiation. The independent variable will be called ``time'',
which corresponds to the main physical applications of the system.

The dependent variables, as well as the matrix elements, may be
complex. We concentrate on real $y$'s and $M$'s because of the
typical physical applications. An analytic extension to complex
values of $y_k, ~k=1,...,n$ (and complex $t$) will be considered
while testing integrability of the equations \eqref{main}.

More general systems, having $A_k\exp(\al_k y_k)$, ~$(k=1,...,n)$,
instead of pure $\exp y_k$\, on the r.h.s. may be transformed into
\eqref{main} by a simple scaling-and-shift transformation of the
dependent variables $y_k\to y'_k$, where
\be\label{equiv}
y'_k=\al_k y_k+\ln A_k-\ln \al_k,~~ M'_{kl}=\al_k\al_l M_{kl}
\ee
which preserves the symmetry of $M$. Some further generalisations,
e.g. to the case where $\forall_k{\al_k<0}$, are also
straightforward.

Also the Hamiltonian model of Adler and van Moerbeke \cite{AM},
where the Hamiltonian
\be\label{AMHam}
\mathcal{H}=\sum_{k=1}^n
p_k^2+\sum_{i=1}^{n+1}\exp\left(\sum_{j=1}^n N_{ij}x_j\right),
\ee
where $N$ is an $(n+1)\times n$ real matrix of rank $n$, may be
transformed into a system whose dynamics equations are
\eqref{main}. This correspondence will be discussed in more detail
while considering the Hamiltonian structure (subsection
\ref{sub-Lagr}) and integrability (section \ref{integr}).

The symmetry of matrix $M$ is relevant for the
Lagrangian-Hamiltonian structure as the inverse of $M$ becomes the
matrix of a quadratic form in $\dot{y}_k$~, k=1,...,n, which plays
the role of the kinetic energy (while $M$ itself is the matrix of
a quadratic form in the momenta -- see subsection \ref{sub-Lagr}).
Otherwise, we can do without this assumption.

 If $M$ may be reduced (by permutation) to diagonal blocks,
we have in fact a few separate systems of equations. Therefore,
only irreducible matrices $M$ will be discussed.

System \eqref{main} emerges from various physical problems, which
are the main motivation for this analysis.
\begin{enumerate}[a)]
\item
The first problem is the Belinski-Khalatnikv-Lifshitz (BKL)
asymptotics of the Einstein equations:

Consider the Einstein equations near the primordial singularity.
Starting from the equations in the synchronous frame of reference
(in which time is the proper time, $t$, at each point); the
authors \cite{BKL1, Ryan} simplify them under the assumption that
(1) the universe is homogeneous but anisotropic, (2) the
anisotropy is indefinitely growing on the approach to the
singularity (like in the classical Kasner solutions
\cite{Kasner}). Then the asymptotic dynamics near the singularity
may be written in terms of the length scales $(a,\,b,\,c)$ in
three principal directions as functions of logarithmic time
$\tau\propto\ln t$ \cite{BKL1, Ryan}. These scale factors satisfy
a set of 3 ODE (the BKL equations), which follow from the
spatial-spatial Einstein equations
\bs\label{BKL-eq}
\be\label{BKL1}\dfrac{d^2 \ln a }{d
\tau^2} = \dfrac{b}{a}- a^2,\quad\dfrac{d^2 \ln b }{d \tau^2} =
a^2 - \dfrac{b}{a} + \dfrac{c}{b},\quad\dfrac{d^2 \ln c }{d
\tau^2} = a^2 - \dfrac{c}{b}\,,\ee while the temporal-temporal
component yields a constraint
\be\label{BKL2}
 \dfrac{d\ln a}{d\tau}
\;\dfrac{d\ln b}{d\tau} + \dfrac{d\ln a}{d\tau}\;\dfrac{d\ln
c}{d\tau} + \dfrac{d\ln b}{d\tau}\;\dfrac{d\ln c}{d\tau} = a^2 +
\dfrac{b}{a} + \dfrac{c}{b}\,,
\ee
\es
whereas the spatial-temporal components of the Einstein equations
merely establish connections between the constants in the
solutions.

In terms of $y_1=\ln (a^2),\, y_2=\ln (b/a), \,y_3=\ln (c/b)$, and
$\tau$ renamed to $t$, the equations take the form of the system
\eqref{main} for $n=3$ with a tridiagonal matrix $M$ (see
\cite{Gold1, Gold2})
\be\label{BKL-M}
M=\left(\ba{ccc}-2&\,2& 0\\2&-2&1\\0&\,1&-2\ea\right).
\ee
The constraint \eqref{BKL2} becomes
\be\label{Ham-3}
\mathcal{H}\colonequals \frac12
\left(\ba{ccc}\dot{y}_1&\dot{y}_2&\dot{y}_3\ea\right)\cdot
M^{-1}\cdot\left(\ba{c}\dot{y}_1\\ \dot{y}_2\\
\dot{y}_3\ea\right)-\left(e^{\,y_1}+e^{\,y_2}+e^{\,y_3}\right)=0.
\ee
We will further see that $\mathcal{H}$ is a first integral of
system \eqref{main}, hence the constraint \eqref{Ham-3} specifies
its value (with $M$ given by \eqref{BKL-M}). Therefore, in spite
of the constraint, this case may be considered within the general
scheme.
\item
The Toda one-dimensional lattice \cite{Toda}. It is a model of a
one-dimensional crystal of \textbf{$n\!+\!1$} nodes (ions, atoms
or molecules) at positions $r_1,...,r_{n+1}$, with $n$
nearest-neighbour interactions. The interactions are defined by
potential
\be\label{Toda-pot}
V(y)=\exp\left(-y\right)+y-1,
\ee
where \textbf{$y$} is a distance between neighbouring nodes. The
interactions cause acoustic oscillations. The model is completely
integrable and has soliton solutions.

The relative positions of the nearest neighbours,
$y_k=r_{k+1}-r_k,~k=1,...,n$, satisfy equation \eqref{main} with a
tridiagonal $n\times n$ matrix $M$, namely
\be\label{Toda2}
 M=\left(\ba{rrrrr}
 2&-1&...&0&0
 \\-1&2&...&0&0\\...&...&...&...&...\\
0&0&...&2&-1\\0&0&...&-1&2 \ea\right).
\ee
\item
There are many problems, which lead to an equation of the form
\eqref{main} for $n=1$, where matrix $M$ reduces to a single
number. In addition to stationary solutions of the one-dimensional
diffusion or Klein-Gordon equation \cite{GrIn}, with exponential
nonlinearity, an interesting example is the Poisson equation for
the electrostatic potential in a non-neutral plasma. Namely,
dynamics of slow phenomena in unmagnetised electron-plasma is
governed by the fluid equations, i.e., \cite{Chen}
\bi\item
Equation of state for the isothermal process
\be\label{eq-state}
p= n_e T,~~T=\mathrm{const},
\ee
where $n_e =$ is the number of electrons per unit volume, $p=$
pressure, $T=$ temperature in energy units.
\item The dynamics equation:
\be\label{eq-dyn}
m n_e[\p_t \vec{v}+(\vec{v}\cdot\nabla \vec{v})]=n_e\, e\,\nabla
\phi - \nabla p,
\ee
where $m$ is the mass of an electron, $\vec{v}=$ is the electron
fluid velocity, $\phi=$ the electrostatic potential.
\item
The left h.s. vanishes in the massless electrons approximation,
$m=0$ (in the slow time scale they react instantly), which allows
for a straightforward solution of system \eqref{eq-state} with
\eqref{eq-dyn}
\be\label{sol m=0}
n_e=n_0\exp(e\phi/T),
\ee
\ei
Equation \eqref{sol m=0} makes a closed system with the field
equation, which (in the absence of magnetic field) is the Poisson
equation. The usual situation includes ion neutralising
background. In 1 dimension
\be\label{plasma}
\p_z^2\phi=e(n_e-n_0)=n_0e[\exp(e\phi/T)-1].
\ee
After rescaling which absorbs the constants $e,~n_0,~T$, we get
the 1-dimensional version of \eqref{main} with an extra term $-1$
for the neutral plasma. This term is absent if we have only
electrons, e.g. an electron beam in the comoving frame of
reference, where we get exactly equation \eqref{main} for $n=1$
(also the $1\times 1$ matrix may be set to $M=1$ under proper
scaling of time) . Its solution is straightforward. Depending on
the assumed positions of the poles, we may write it (for general
$M$) as e.g.,
\be\label{n=1}
y=\ln\left\{\frac{2k^2}{M}\sec^2\left[k(t-t_0)\right]\right\}
\text{ or e.g., }
y=\ln\left\{\frac{2k^2}{M}\csch^2\left[k(t-t_0)\right]\right\}
\ee
with arbitrary $k$ and $t_0$.
\end{enumerate}
In Section 2, we discuss the basic properties of system
\eqref{main}, valid for any number of equations $n$. These
properties include the Lagrangian and Hamiltonian structure,
symmetries and conservation laws, a family of exact solutions, and
instability of the exact solutions to small perturbations. In
Section 3, the question of integrability is addressed. The
analysis, using similar results of \cite{AM} shows that the only
systems which are integrable are similar to the Toda chain
\cite{Toda}.

\section{General properties of the system \eqref{main}}
In this section, we discuss \eqref{main} as a Lagrangian and
Hamiltonian system, demonstrating its structure and correspondence
with the system discussed in \cite{AM}. The two Lie symmetries:
translation of the independent variable and a scaling of this
variable combined with a common shift of the dependent variables
are shown. Then the exact solution is found, which proves to be
the self-similar solution with respect to the latter symmetry.
Finally, the first-order solution for small perturbations of the
initial conditions is calculated; the ratio of the perturbations
to the perturbed quantities is shown to increase in time $t$ as
$t^{1/2}$.
\subsection{Lagrangian and Hamiltonian structure}\label{sub-Lagr}
\begin{theorem}
System \eqref{main} consists of the Lagrange equations for a
Lagrangian
\be\label{Lagr}
\mathcal{L}\colonequals\frac12
\sum_{k,l=1}^n\dot{y}_k{M^{-1}}_{kl}\dot{y}_l+\sum_{k=1}^n\exp
y_k.
\ee
\end{theorem}
\begin{pf}
The Euler-Lagrange equations
$\frac{d}{dt}\frac{\p\mc{L}}{\p\dot{y}_k}-\frac{\p\mc{L}}{\p
y_k}=0$ become our system \eqref{main} if we make use of the fact
that an inverse of a symmetric matrix is also a symmetric
matrix.\qed
\end{pf}
The Legendre transformation to the canonical momenta is
straightforward
\be\label{momentum}
p_k=\sum_{l=1}^n {M^{-1}}_{kl}\dot{y}_l,\quad
\dot{y}_k=\sum_{l=1}^n M_{kl}p_l.
\ee
Absence of explicit time dependence in the Lagrangian \eqref{Lagr}
infers conservation of the Hamiltonian \eqref{Ham}, whether it is
a quadratic form expressed in terms of canonical momenta
\be\label{Ham}
\mathcal{H}\colonequals\frac12\sum_{k,l=1}^n p_k M_{kl}
p_l-\sum_{k=1}^n\exp y_k
\ee
or in terms of time derivatives $\dot{y}_k,~k=1,...,n$, with
$M^{-1}$ being the matrix of the quadratic form as in eq.
\eqref{Ham-3}. 
\begin{remark}
In the notation of \cite{AM}, the Hamilton equations corresponding
to their Hamiltonian \eqref{AMHam} lead to \eqref{main} upon
substitution
\be\label{AM2main}
y_k=\sum_{l=1}^n \bar{N}_{kl}x_l, ~~M=\bar{N}^T\cdot \bar{N},
\ee
where $\bar{N}$ is the square matrix obtained from their
$(n+1)\times n$ matrix $N$ by dropping its last row.
\end{remark}
\subsection{Symmetries}
Equations \eqref{main} are obviously symmetric with respect to
time reversal $t\rightleftarrows -t$ (as they contain only second
time derivatives) and time shift $t\rightarrow t-t_0$ (as
autonomous equations). A less obvious but easily seen symmetry
combines scaling of time with a shift in the dependent variables
\be\label{scaling}
t\rightarrow \al t,~~y_k\rightarrow y_k-2\ln\al,~~\al>0,
\ee
with the same $\al$ for all $k=1,...,n$. The last symmetry
corresponds to a vector field
\be\label{Lievector}
t\frac{\p}{\p t}-2\sum_{k=1}^n \frac{\p}{\p y_k},
\ee
to which the prolongations do not add any terms.
\subsection{The exact solutions}
In \cite{GP}, an exact solution was first found, for the special
case of the BKL scenario. It may easily be generalised to
solutions of \eqref{main}. Namely
\begin{theorem}
For almost all real symmetric matrices $M$,
\be\label{exact}
y_k=\ln\frac{a_k}{t^2},~k=1,...n,~\mathrm{where}~a_k\colonequals
2\sum\limits_{l=1}^n {M^{-1}}_{k\,l},
\ee
is an exact solution of \eqref{main}.
\end{theorem}
\begin{pf}
Consider both sides of equation \eqref{main} left-multiplied by
matrix $M^{-1}$. Substitution of \eqref{exact} to its lhs yields a
column vector with elements
\be\label{ak}
\frac2{t^2}\sum_{l=1}^n{M^{-1}}_{kl}= \frac{a_k}{t^2},~k=1,...,n.
\ee
The substitution of \eqref{exact} to the rhs of this equation
yields also a column with elements $\exp\ln (a_k/t^2)=a_k/t^2$.
Multiplying both resulting sides by $M$ we get the required
equality of both sides in \eqref{exact}. The only exception is the
situation where the logarithm of $a_k$ does not exist, which is
the situation described in Remark~\ref{exc}.\qed
\end{pf}
\begin{remark}\label{exc}
The exact solution does not exist for a subset of measure zero in
the space of $n\times n$ real symmetric matrices $M$ (the measure
may be e.g. induced by the Lebesque measure on
$\mathbb{R}^{n(n+1)/2}$). Namely, the exact solution does not
exist for the matrices which satisfy
\be\label{exception}
\sum_{l=1}^n{M^{-1}}_{kl}=0
\ee
for at least one $k\in \{1,...,n\}$.
\end{remark}
\begin{remark}
The exact solution has a second order pole at $t=0$. Due to the
translational symmetry, it may be shifted to any
$t_0\in\mathbb{R}$. The value of $t_0$ depends on the initial
conditions; thus $t_0$ is a movable pole in the sense of
Painlev\'e \cite{Conte2}.
\end{remark}
\begin{remark}
The exact solution \eqref{exact} is the self-similar solution with
respect to the scaling symmetry \eqref{scaling}. This property is
apparent if we directly apply the symmetry transformation
\eqref{scaling} to the solution \eqref{exact}.
\end{remark}
\begin{remark}
For $n=1$, solution \eqref{exact} is the limit case of the second
solution of \eqref{n=1}, as $k\to 0$.
\end{remark}
\begin{remark}
The exact solution \eqref{exact} (whether the pole is at $t=0$ or
at some other $t_0$) corresponds to $\mc{H}=0$ as in \eqref{Ham-3}
(a constraint typical for the models of gravity like the BKL),
which may be easily seen by substitution of this solution to
\eqref{Ham} with \eqref{momentum}, or \eqref{Ham-3}. A
two-parameter class of exact solutions may be found if we do not
impose this constraint.
\end{remark}
In fact, we have
\begin{theorem}
System \eqref{main} has a two-parameter class of exact solutions
\be\label{generalised}
y_k=\ln\left[b^2 a_k\csch^2b(t-t_0)\right],~k=1,...n;~\mathrm{as
before}~a_k\colonequals 2\sum\limits_{l=1}^n {M^{-1}}_{k\,l},
\ee
\end{theorem}
The proof is identical (mutatis mutandis) to that for solution
\eqref{exact}. This type of solution was mentioned in the context
of the BKL scenario in \cite{Conte3}.
\begin{remark}
The one-parameter solution \eqref{exact} is a straightforward
limit of the solution \eqref{generalised} for $b\to 0$.
\end{remark}
\subsection{Stability to small perturbations}\label{stability}
In the stability analysis, as well as in the integrability tests,
a special role is played by a matrix $R$
\be\label{R}
R_{kl}\colonequals a_k M_{kl},~k=1,...,n, \text{ or } R =
\diag(a_1,...,a_n)\cdot M,
\ee
where $a_k$ is defined in \eqref{exact}. The second formulation of
\eqref{R} shows that (with the exception of \eqref{exception}),
this matrix is nonsingular and diagonalisable.

The convenient variables for the stability analysis of the exact
solution \eqref{exact} are
\be\label{q-var}
q_k=\exp{y_k}, ~\text{with the exact solution}~\,
q_k=a_k/t^2,~\,k=1,...,n,
\ee
because equations \eqref{main} may be transformed into a system of
polynomial equations in these variables.
\be\label{polyn}
q_k\ddot{q}_k-\dot{q}_k^2=q_k^2\sum_{l=1}^n M_{kl}q_l
\ee
We first prove instability of the one-parameter exact solution
\eqref{exact}. Assume that a small perturbation $\ep\al_k(0),
~k=1,...,n,~\ep>0$ is added to each $q_k$ at $t=0$. Consider the
dynamics of $\al_k(t)$ by solving the perturbed equation
\eqref{main} expressed in terms of the $q_k$ and linearised to
\ist order in $\ep$. A straightforward calculation leads to
\be\label{1st eq}
\left(t^2
\al_k\right)^{\scriptscriptstyle\bullet\bullet}=\sum_{l=1}^n
R_{kl}\al_l,~k=1,...,n.
\ee
The system is equidimensional (aka Euler or Cauchy-Euler), i.e.,
the coefficients at derivatives of $\al_k$ are power functions
$\propto \,t^p$, where $p$ matches the order of the derivative. A
well-known theorem states \cite{Ince} that the general solution of
system \eqref{1st eq} is a polynomial, namely it is an arbitrary
linear combination of power functions $\al_k(t)=c_{kr} t^r$, whose
exponents $r$ may be obtained by substitution of this solution to
the ODE. The substitution yields an algebraic equation for $r$. If
the algebraic equation has a root of multiplicity $m$, then terms
of the form $\propto\,t^r(\ln t)^q,~q=1,...,m\!-\!1$ occur in the
general solution, in addition to the power-function terms.

Substitution of $t^r$ into \eqref{1st eq} leads to a system of
algebraic equations
\be\label{algeq}
(r+2)(r+1)\,c_{kr}=\sum_{l=1}^n R_{kl}c_{lr}.
\ee
It follows from \eqref{algeq} that $(r+2)(r+1)$ must be one of the
eigenvalues $\lam_1,...,\lam_n$ of matrix $R$ (the enumeration
includes repetition if some of the eigenvalues are equal). Hence,
we have $2n$ solutions for $r$ if they are counted with their
multiplicity. From eq. \eqref{algeq}, it follows
\begin{lemma}\label{lemma}
For each $\lam_k$, at least one of the solutions of
\be\label{quadreq}
(r+2)(r+1)=\lam_k
\ee
satisfies $\Re{r}\ge -3/2$.
\end{lemma}
\begin{pf}
The thesis of the lemma directly follows from Vieta's formula: If
$r_1$ and $r_2$ are solutions (unequal or equal to each other) of
equation \eqref{quadreq}, then $r_1+r_2=-3$.\qed
\end{pf}
This implies the instability of the exact solution \eqref{exact}.
Namely,
\begin{theorem}\label{th3}
There are initial conditions of the perturbation
$\al_1(0),...,\al_n(0),\dot{\al}_1(0),...,\dot{\al}_n(0)$, having
nonzero measure in $\mathbb{R}^{2n}$, for which the ratio of the
perturbation $\al(t)$ to the perturbed solution \eqref{q-var}
increases at least as $t^{1/2}$ or oscillates with amplitude
increasing at least as $t^{1/2}$.
\end{theorem}
\begin{pf}
Let $r_{max}$ be the root with the greatest real part of all
$r_1,...,r_n$. Consider the term $c_{kr}t^r$ for $r=r_{max}$. The
set with $c_{kr}= 0$ is of measure zero in $\R^{2n}$, hence the
set with $c_{kr}\ne 0$ and all possible $c_r$ has nonzero measure.
If the root $r=r_{max}$ is real, then $r\ge 3/2$ and the ratio of
this term to corresponding term of the exact solution satisfies
\be\label{amp}
\lvert c_{kr} t^r/\left(a_k t^{-2}\right)\rvert\ge \lvert c_{kr}
/a_k\rvert t^{1/2},
\ee
i.e., it increases as $t^{1/2}$. For large $t$ this term is
greater than the other terms. If $r_{max}$ is complex,
$r_{max}=x+y i,~\,x\ge -3/2$, then $t^r=t^x\exp(i y\ln t)$, which
describes oscillations (of diminishing frequency). Then the
asymptotic \eqref{amp} concerns the amplitude, $c_{kr}t^x$ rather
than the whole term. In both cases,
\be\label{inst}
\sup_{t< T}\lvert\al_k(t)/q_k(t)\rvert\to\infty \text{ as
}T\to\infty
\ee
for all $k$ having nonzero coefficients $c_{kr}$ in the initial
condition. This means instability of each $q_k$ (and consequently
$y_k=\ln q_k$) to small perturbations. \qed
\end{pf}
\begin{remark}
If the logarithmic time $\tau=\ln t$ is used as the independent
variable (which is also a well-known method of solving the
equidimensional equations), then $\exp(-2\tau)$ is the rate of
convergence of the exact solution, while $\exp(-3\tau/2)$ is the
corresponding rate for the perturbation. Hence, $-2$ and $-3/2$
play the role of the Lyapunov exponents, while the positive sign
of their difference, $\Delta\ge 1/2$, shows that the the exact
solution \eqref{exact} behaves as a repeller.
\end{remark}
\begin{corollary}
For the two-parameter exact solution \eqref{generalised}, the
above argument describes the onset of the linear instability. In
this case, the perturbation $(\al_1,...,\al_n)$ satisfies a
generalisation of \eqref{1st eq}, namely
\be\label{pert-gen}
b^{-2}\left[\al_k\sinh^2 \left(bt
\right)\right]^{\scriptscriptstyle\bullet\bullet}=\sum_{l=1}^n
R_{kl}\al_l,~k=1,...,n.
\ee
In the first order in $b$, \eqref{pert-gen} turns into \eqref{1st
eq}. The same applies to the first order in $t$, which shows that
the initial rate of growth, for some of the ratios $\al_k/q_k$,
would be at least as $t^{1/2}$.
\end{corollary}
\section{Integrability}\label{integr}
Integrability of a similar model has been analysed in detail in a
classical paper by Adler and van Moerbeke \cite{AM}. Rather than
calculate the criteria of integrability from scratch, we recall
their result and apply it to our case, using transformation
\eqref{AM2main}.

Authors of \cite{AM} combine the Painlev\'e test \cite{ARS,
Conte2} with Lie algebra technique for determining the conditions
of integrability.

Typically, the test is used to determine the necessary conditions
for the Painlev\'e property. To remind the reader, the test relies
on substitution of a hypothetic solution, in the form of a Laurent
series about a pole, to the original equation expressed in a
polynomial form. After finding the dominant term(s), the
subsequent ones are calculated from algebraic recurrence
relations. The orders where the the relations become
underdetermined or contradictory (``resonances'' or ``indices'')
are calculated from the indicial equation. The underdetermined
provide first integrals, while existence of a contradictory
relation means that the system is non-Painlev\'e \cite{Conte2}.

In the usual applications of the test to the ODE having the
Painlev\'e property, the test should provide as many first
integrals as the total order of the system, which is $2n$ in our
case. However,
\begin{remark}
Obtaining $2n$ first integrals from the recurrence relations is
not possible for our system \eqref{main}, because the indicial
equation at $r$'th order of the expansion reads
\be\label{indicial}
r(r-1)-\lam_k=0,~~k=1,...,n,
\ee
where $\lam$ is an eigenvalue of the $R$ matrix defined in
\eqref{R}. This equation has a half of the solutions with their
real part less or equal to $-\tfrac14$ (by an argument similar to
Lemma \ref{lemma}), whence the relations may provide at most $n$
first integrals in the system of order $2n$. Although an extension
of the Painlev\'e test to negative indices exists \cite{CFP}, only
its negative answer would be conclusive, while we are looking for
integrable cases.
\end{remark}
Luckily, a stronger result, namely integrability by quadratures,
may be achieved with only $n$ first integrals, if the system is
Hamiltonian and the first integrals are global and in involution
\cite{Arnold1}. This is the goal of \cite{AM}.

The model of \cite{AM} starts with Hamiltonian \eqref{AMHam}. The
authors of prove that the system is integrable if and only if the
ratio between the following elements of matrix $N$ (defining the
Hamiltonian \eqref{AMHam}) satisfies
\be\label{ratio}
2(N\cdot N^T)_{ij}/(N\cdot N^T)_{jj}\text{ is a nonpositive
integer for all $i\ne j$}.
\ee
This is a property of the Toda systems, described by potential
\eqref{Toda-pot}, generalised by inclusion of the non-nearest
neighbour interactions \cite{AM}.

Translating the result of \cite{AM} to our systems (see
\eqref{AM2main}), the criterion, limited to $N$ without its
$(n+1)$-th row, simply applies to our symmetric matrix $M
=\bar{N}^T\cdot \bar{N}=\bar{N}\cdot \bar{N}^T$, as
\be\label{AM-crit}
\forall_{i\ne j}M_{ij}/M_{jj}\in \mathbb{Z}^-,
\ee
However, an analogous condition has to be imposed on the last row
of the $(n+1)\times n$ matrix $N$, and the row should be
constructed in such a way that $N^T$ had a nonzero null vector
$p=(p_1,...,p_n,1)^T\equalscolon (\bar{p},1)^T$ (this symbol $p$
of \cite{AM} has no connection with the momentum of
\eqref{momentum}). In fact, it is sufficient that $N\cdot N^T$ had
such a null vector.

The construction of the appropriate extended matrix $N\cdot N^T$
(which the authors denote by $M$, while our symbol $M$ corresponds
to their $\bar{M}$) relies on completing our $M$ with the
$(n+1)$-th row $-(M\cdot \bar{p})^T,\bar{p}\cdot M\cdot\bar{p}^T$,
and the $(n+1)$-th column which is its transposition. Then $p$ is
automatically the required null vector. This construction is
always possible, but not always would the elements of the extra
row and column satisfy the condition \eqref{AM-crit}. A
counterexample is the BKL scenario, whose matrix $M$ \eqref{BKL-M}
apparently satisfies \eqref{AM-crit}, but its extension does not,
whatever be the null vector. Indeed, the Painlev\'e test shows
that the BKL model is not integrable \cite{GP}.

\section{Conclusions}
We have analysed a system of $n$ equations \eqref{main} with
exponential nonlinearity. The system proves to have various
applications, from behaviour of the universe close to the
cosmological singularity \eqref{BKL1}, \eqref{BKL2} through
crystals \eqref{Toda2}, to plasma physics, \eqref{plasma}. For
this system of ODE's, we have found a two-parameter family of
exact solutions . The solutions prove to be unstable to small
perturbations; more precisely, the ratio of the perturbation to
the perturbed quantity increases as $t^{1/2}$ or oscillates with
amplitude increasing at this rate. The known criterion of
\cite{AM} on integrability of similar models \eqref{AM-crit} may
be transformed to describe our systems by a simple transformation.
The criterion does not apply to cases with $n=1$; however these
cases may be explicitly solved, yielding solution \eqref{n=1}.















\end{document}